\documentclass[aps,prl,reprint,showpacs,floatfix]{revtex4-2}
\usepackage{amsmath,amssymb,graphicx}

\begin{document}


\title{Entanglement Before Spacetime in Quantum-Gravity-Induced Interactions}

\author{Hollis Williams}
\affiliation{Department of Mathematics and Statistics, University of Exeter, Exeter EX4 4QF, UK}

\begin{abstract}Quantum-gravity-induced entanglement of massive systems (QGEM) is commonly approximated in the nonrelativistic static limit by a Newtonian interaction between spatially separated masses. In this work, we reformulate the gravitationally mediated interaction phase in a conformally invariant twistor framework in which no notion of spacetime distance is assumed. We show that the bilocal phase responsible for entanglement generation remains well-defined and non-factorizable even in the absence of spacetime geometry. The familiar Newtonian $1/r$ phase, relevant for QGEM protocols, arises only after the conformal invariance is broken by introducing the infinity twistor, which selects a particular spacetime representation of the underlying bilocal quantum interaction. Our results isolate the genuinely quantum content of QGEM protocols and clarify the contingent role played by spacetime geometry in mediating entanglement.
\end{abstract}

\maketitle

\section{Introduction}

 \noindent
Quantum-gravity-induced entanglement of massive systems (QGEM) has been proposed as a table-top probe of the quantum nature of gravity, based on the observation that purely classical channels cannot generate entanglement between spatially separated systems prepared in a quantum superposition \cite{bose, carney}. In the simplest implementations, two masses are placed in spatial superposition and interact gravitationally for a fixed duration. The resulting entanglement is attributed to a relative phase generated by the gravitational interaction.  This is usually by a Newtonian potential in the nonrelativistic static limit \cite{vedral}. The operational quantity governing entanglement generation is then the phase \( \Phi \sim G m_A m_B T / \hbar r \), and observation of entanglement is interpreted as evidence that gravity cannot be described by a purely classical mediator \cite{vedral2}.

A substantial body of work has explored refinements and extensions of this basic idea, including schemes with improved robustness to decoherence, many-body generalizations, Casimir screening effects, and formulations in curved or higher-dimensional spacetimes \cite{nguyen, schut1, ghosal, liu, kamp, zhang, zhang2, elahi, feng}. In nearly all such analyses, the gravitational interaction phase is computed from spacetime-dependent quantities (a Newtonian potential or a proper time difference) which are treated as the fundamental inputs into the protocol.  Although this approach is natural from the perspective of nonrelativistic gravity, it leaves implicit the extent to which spacetime itself is essential for the mechanism of entanglement generation.

From an effective field theory viewpoint, the gravitational interaction between two localized masses arises when one integrates out a massless mediator, yielding a bilocal influence functional which couples their worldlines \cite{feynman, dono}. The resulting unitary evolution is characterized by a non-factorizable phase functional whose existence ensures the possibility of entanglement generation.  Crucially, this bilocal structure is logically distinct from any particular geometric representation of the interaction, such as a distance-dependent spacetime metric. Although QGEM protocols rule out purely classical channels, focusing exclusively on spacetime-based descriptions can obscure the more primitive role played by this non-factorizable bilocal phase \cite{hall, rovelli}. This motivates the search for a formulation in which the bilocal interaction can be defined independently of a spacetime metric, whilst still reproducing the familiar Newtonian limit when additional structure is introduced.

In this work, we make this distinction explicit by reformulating the bilocal gravitational interaction in a twistor framework, in which spacetime distance and metric structure are not assumed a priori \cite{penrose, held}. Twistor theory provides a natural language for describing conformal structure and propagation of massless particles, allowing the interaction phase to be defined directly in terms of invariant relations between worldlines rather than spacetime separations.  In this formulation, the bilocal phase responsible for entanglement generation remains well-defined and non-factorizable even in the absence of a metric.  We show that the Newtonian \(1/r\) interaction phase relevant for QGEM protocols emerges only after conformal invariance is broken via introduction of an infinity twistor, which selects a particular spacetime interpretation for the underlying bilocal quantum interaction \cite{rindler}. Geometry therefore enters not as a prerequisite for entanglement generation, but as a representational structure imposed on a quantum channel which already exists prior to this choice.

Our results clarify the operational content of QGEM-type experiments by isolating the genuinely quantum ingredient which they probe (existence of a non-factorizable bilocal interaction) from the geometric assumptions used to describe it. In this sense, gravitationally mediated entanglement occurs ``before spacetime'': the interaction responsible for entanglement is defined prior to the introduction of a distance measure or a spacetime metric, with spacetime locality arising only as an emergent description of the underlying quantum structure.  Before conformal symmetry is broken, the bilocal phase governing QGEM interactions is necessarily scale-free, implying that no metric notion of spatial separation can be defined at this level.

\section{Bilocal interaction}

\noindent
At energies well below the Planck scale, the gravitational interaction between two localized massive systems $A$ and $B$ can be described within an effective field theory by integrating out the metric perturbations. This yields a bilocal interaction between their stress tensors 

\begin{equation} S_{\text{int}} = \frac{1}{2}  \int d^4 x \: d^4 x' \: T_{A}^{\mu \nu} (x) D_{\mu \nu \rho \sigma} T_{B}^{\rho \sigma} (x') , \end{equation}

\noindent
where $D_{\mu \nu \rho \sigma}$ is the graviton propagator.  For a point particle following a worldline, the stress tensor may be written as 

\begin{equation}  T^{\mu \nu} (x) = m \int d\tau \: \frac{u^{\mu} u^{\nu}}{\sqrt{-g}} \delta^{(4)} (x - x(\tau)) ,     \end{equation}

\noindent
where $m$ and $u^{\mu}$ are the mass and four-velocity of the particle, respectively.  Substituting this back in, we obtain

\begin{equation}  S_{\text{int}} = \frac{1}{2} \int d \tau \: d \tau' \: m_A m_B u_A^{\mu} u_B^{\nu} D_{\mu \nu \rho \sigma} ( x_A(\tau), x_B(\tau')) u_B^{\rho} u_B^{ \sigma}.\end{equation}

\noindent
Note that this action is still fully relativistic. 


To simplify in this work, we will consider a scalar version of this interaction which is obtained by suppressing Lorentz indices and replacing the graviton propagator with the Green's function of a massless mediator $G(x,x')$: 

\begin{equation}
S_{\mathrm{int}}
=
\frac{1}{2}
\int d\tau \, d\tau'\;
m_A m_B\,
G\!\left(x_A(\tau),x_B(\tau')\right),
\label{eq:bilocal_action}
\end{equation}
 
\noindent
The Green's function corresponds schematically to the graviton propagator contracted with the conserved worldline currents, since the scalar structure is sufficient for present purposes.  The contribution of the interaction \eqref{eq:bilocal_action} to the unitary evolution of the joint system is given by the phase
\begin{equation}
\Phi_{AB}
=
\frac{1}{\hbar}
S_{\mathrm{int}}.
\label{eq:bilocal_phase}
\end{equation}
Whenever the kernel $G(x,x')$ is nontrivial, the resulting unitary operator
\begin{equation}
U_{AB} = \exp\!\left(i \Phi_{AB}\right)
\end{equation}
is non-factorizable and can therefore generate entanglement between the two massive systems when they are prepared in a spatial superposition.

As noted above, one can connect with QGEM-type protocols by considering the nonrelativistic static limit in which the two masses remain approximately at rest at a fixed spatial separation $r = |\mathbf{x}_A - \mathbf{x}_B|$. In this limit, the dominant contribution to the Green's function is instantaneous and one finds
\begin{equation}
G\!\left(x_A(\tau),x_B(\tau')\right)
\;\longrightarrow\;
-\,G\,\frac{\delta(t-t')}{r}.
\end{equation}
Substituting this into Eq.~\eqref{eq:bilocal_action} yields
\begin{equation}
\Phi_{AB}
\;\longrightarrow\;
-\frac{G m_A m_B}{\hbar}
\int dt \, \frac{1}{r}.
\label{eq:newtonian_phase}
\end{equation}
For an interaction time $T$, this gives a total phase
\begin{equation}
\Phi_{AB}
\sim
\frac{G m_A m_B T}{\hbar r},
\label{eq:newtonian_phase2}
\end{equation}
in agreement with the expressions appearing in \cite{bose, vedral}.

We emphasize that although equation~\eqref{eq:newtonian_phase2} is usually assumed in QGEM analysis, it actually arises from two distinct steps.  Firstly, one assumes the existence of a bilocal interaction phase \eqref{eq:bilocal_phase} determined by the propagation kernel $G(x,x')$, and secondly, this phase is identified with a Newtonian potential proportional to $1/r$ via a static equal-time limit. In the following section, we will reformulate the bilocal phase \eqref{eq:bilocal_phase} in a representation which does not assume a spacetime notion of distance, isolating which features of QGEM-type protocols are intrinsic to the interaction and which depend on additional geometric structure.

\section{Twistor formulation}

\noindent
The scalar bilocal interaction \eqref{eq:bilocal_action} depends on the propagation kernel of a massless mediator between two massive worldlines.  In order to remove dependence on spacetime points and make manifest when it becomes necessary to assume spacetime structure, we will reformulate the interaction with twistor formalism \cite{penrose}.  We therefore replace the spacetime Green's function $G(x,x')$ with a twistor space kernel $K(Z,Z')$, where $Z$ and $Z'$ denote twistor variables. The kernel $K(Z,Z')$ is assumed to satisfy the following properties: (i) it represents massless propagation; (ii) it encodes causal correlations; and (iii) when we choose appropriate additional structure, its Penrose transform reproduces the standard spacetime Green's function.

With this replacement, the bilocal interaction phase may be written schematically as
\begin{equation}
\Phi_{AB}
=
\frac{m_A m_B}{2\hbar}
\int d\tau \, d\tau'\;
K\!\left(
\mathcal{Z}_A(\tau), \mathcal{Z}_B(\tau')
\right),
\label{eq:twistor_bilocal_phase}
\end{equation}
where $\mathcal{Z}_A(\tau)$ and $\mathcal{Z}_B(\tau')$ denote the twistor data encoding the two timelike worldlines (to be defined shortly).  In twistor theory, a single twistor encodes the geometry of a null ray in spacetime and therefore provides a natural description for massless particles. Since QGEM protocols involve massive particles which follow timelike curves, a single twistor is insufficient to represent the relevant degrees of freedom.  Instead, a standard way to encode a massive particle in twistor language is via a pair of twistors subject to appropriate constraints, known as a bitwistor. Geometrically, a bitwistor corresponds to a timelike direction in spacetime along with an associated mass scale.  We will briefly summarize the construction and detailed reviews may be found in \cite{penrose}.

A simple bitwistor for particles $A$ and $B$ may be written as 
\begin{equation}
X^{\alpha\beta}_A = Z_A^{[\alpha} W_A^{\beta]}, 
\qquad
X^{\alpha\beta}_B = Z_B^{[\alpha} W_B^{\beta]},
\end{equation}
where $Z,W \in \mathbb{CP}^3$ and the simplicity constraint
$X^{[\alpha\beta}X^{\gamma\delta]}=0$ ensures that each bitwistor corresponds to a timelike worldline. The description is invariant under independent $GL(2,\mathbb{C})$ transformations acting on $(Z_A,W_A)$ and $(Z_B,W_B)$.  A timelike worldline is obtained by considering a
one-parameter family of such bitwistors $\mathcal{Z}(\tau)$.  In this formulation, the spacetime trajectories $x_A(\tau)$ and $x_B(\tau')$ appearing in the bilocal interaction are replaced by one-parameter families of bitwistors $\mathcal{Z}_A(\tau)$ and $\mathcal{Z}_B(\tau')$.  The gravitationally induced bilocal phase may therefore be viewed as a functional of two bitwistor worldlines coupled via a massless propagation kernel, without assuming a spacetime metric or a notion of distance.


\section{Invariant structures in twistor space}

\noindent 
To summarize, we describe the interaction between the two worldlines in twistor space by a bilocal kernel $K(X,X')$, which is an abstract scalar function defined on bitwistor space.  Its functional form is initially unspecified, but it is constrained by conformal invariance, projective rescalings, and $GL(2,\mathbb{C})$ transformations on each bitwistor.  We next look to classify all the scalars which may appear in the interaction phase after imposing these symmetries.  The two bitwistors may be contracted over to form an independent conformal invariant

\begin{equation}
I_{AB} := X_A^{\alpha\beta} X_{B\,\alpha\beta}.
\end{equation}

\noindent
This quantity is dimensionless and invariant under all symmetries listed above. It characterizes the relative incidence of the two worldlines but does not encode any notion of spatial separation.  Another possibility for obtaining invariants is to evaluate the scalar kernel on the twistor components of the two bitwistors, producing quantities such as $K(Z_A,Z_B)$ and $K(Z_A,W_B)$. However, none of these are invariant under $GL(2,\mathbb{C})$. The unique invariant combination is the determinant
\begin{equation}
\mathcal{K}_{AB}
=
\det
\begin{pmatrix}
K(Z_A,Z_B) & K(Z_A,W_B) \\
K(W_A,Z_B) & K(W_A,W_B)
\end{pmatrix}.
\end{equation}
This object is invariant under independent $GL(2,\mathbb{C})$ transformations on each bitwistor, under projective rescalings, and under the action of the conformal group $SL(4,\mathbb{C})$.

It follows that, prior to any breaking of conformal invariance, the interaction phase can depend on the bitwistor worldlines only through scalar invariants constructed from the corresponding bitwistors.  Once the appropriate symmetries are enforced, the most general bilocal phase must take the form
\begin{equation}
\Phi_{AB}
=
\int d\tau \, d\tau'\;
F\!\left(
I_{AB},
\mathcal{K}_{AB}
\right),
\end{equation}
where $I_{AB}$ and $\mathcal{K}_{AB}$ are the scalar invariants defined above.  Notably, no notion of spatial distance or temporal separation appears at this stage, and no dimensionful invariant can be constructed.

Since the Newtonian interaction phase scales as $1/r$, the above analysis shows that such a dependence cannot arise within a fully conformally invariant twistor description.  Any recovery of a Newtonian $1/r$ phase therefore requires the introduction of additional structure that breaks conformal invariance, such as an infinity twistor.  To illustrate these concepts concretely, we now fix the twistor kernel and focus on the previously introduced free massless scalar field.  We then compute the bilocal phase explicitly along bitwistor worldlines to connect with the Newtonian limit and QGEM protocols.

\section{Scalar mediation}

\noindent
The interaction between two massive worldlines mediated by a free massless scalar field is sufficient to capture the structure of the QGEM interaction phase, since this phase only depends on the existence of a long-range massless mediator which generates a non-factorizable phase.  The spin-2 nature of the graviton only modifies tensor structure and does not affect the operational generation of entanglement in the static limit.  

As stated above, in a worldline EFT description such an interaction gives rise to a bilocal phase proportional to the massless Green’s function evaluated on the two worldlines.  In a twistor formulation, a free massless scalar field is represented by a scalar kernel \(K(Z,Z')\) on projective twistor space which is homogeneous of degree \((-1,-1)\) in its arguments. When pulled back to spacetime, this kernel reproduces the usual massless propagator. At this stage, however, no spacetime interpretation is assumed: the kernel is only required to respect the symmetries outlined above.

In a fully conformally invariant setting, there is no canonical two-point scalar invariant on twistor space. In particular, without introducing additional structure, there is no distinguished bilinear form with which to contract two twistors. As a result, the scalar kernel \(K(Z,Z')\) is either not well defined or ambiguous up to arbitrary choices.  The corresponding bitwistor invariant \(\mathcal{K}_{AB}\) which we constructed in the previous section is therefore either trivial or non-unique. This implies that no nontrivial bilocal interaction phase can be defined with a strictly conformally invariant twistor framework.

To proceed, we therefore introduce an infinity twistor \(I_{\alpha\beta}\), which breaks conformal invariance and endows twistor space with a notion of scale.  This twistor \(I_{\alpha\beta}\) picks a preferred conformal frame and allows the identification of twistor incidence relations with spacetime points. In particular, for twistors \(Z^\alpha = (\omega^A,\pi_{A'})\) satisfying the incidence relation
\begin{equation}
\omega^A = i x^{AA'} \pi_{A'},
\end{equation}
the contraction
\begin{equation}
\langle Z Z' \rangle = I_{\alpha\beta} Z^\alpha Z'^\beta
\end{equation}
reduces up to a factor to the square of the spacetime separation between the corresponding points up to an overall normalization.  This allows us to define a scalar twistor propagator
\begin{equation}
K(Z,Z') = \frac{1}{\langle Z Z' \rangle},
\end{equation}
which has the correct homogeneity properties and reduces under the Penrose transform to the standard massless Green’s function in spacetime. 

Substituting back into equation (13), we obtain 

\begin{equation}
 \mathcal{K}_{AB} = \langle Z_A Z_B \rangle^{-1}  \langle W_A W_B \rangle^{-1} -  \langle Z_A W_B \rangle^{-1}  \langle W_A Z_B \rangle^{-1}    .
\end{equation}

\noindent
Since $Z_A$ and $W_A$ span the same bitwistor plane, spinor normalizations cancel out and we find that

\begin{equation}
  \mathcal{K}_{AB} \propto  \frac{1}{(x_A - x_B)^2}. 
\end{equation}

\noindent
The determinant removes twistor gauge dependencies and becomes a scalar function of the invariant spacetime distance between two worldlines.  This expression is still fully relativistic and Lorentz invariant.

\section{Static limit}

\noindent
In the regime relevant to QGEM protocols, the trajectories are nonrelativistic and well separated, so that
\begin{equation}
(x_A(\tau) - x_B(\tau'))^2
=
-(t - t')^2 + |\mathbf{x}_A(t) - \mathbf{x}_B(t')|^2 .
\end{equation}
Evaluating the bilocal phase in the static limit and integrating over coordinate time yields
\begin{equation}
\Phi_{AB}
=
-\frac{G m_A m_B}{\hbar}
\int dt \, \frac{1}{r(t)},
\end{equation}
where $r(t)=|\mathbf{x}_A(t)-\mathbf{x}_B(t)|$, reproducing the Newtonian interaction phase employed in QGEM analyses.  This result arises only after conformal symmetry is broken by the introduction of the infinity twistor (or an equivalent choice).  Prior to this, no notion of spatial distance or Newtonian interaction is available.

\section{Discussion}

\noindent
The analysis carried out here shows that QGEM-type protocols do not fundamentally test spacetime locality, but rather the existence of a non-factorizable bilocal interaction between two massive quantum systems. This interaction is naturally characterized by a bilocal phase functional, which can be formulated without reference to a spacetime metric or spatial separation.  In a twistor formulation, the bilocal phase is defined directly in terms of invariant structures associated with the two worldlines.  Crucially, this phase (and therefore the possibility of entanglement generation) exists before any identification of spacetime distance and persists in fully relativistic formulations of QGEM protocols, where entanglement is mediated by gravitational interactions without invoking a Newtonian static limit \cite{higgins}.  The familiar Newtonian behaviour arises only after additional geometric structure is introduced, breaking conformal invariance and choosing a particular spacetime representation of the underlying interaction.

From an information-theoretic perspective, the bilocal phase defines a non-factorizable quantum channel between the two massive systems. Entanglement generation in QGEM protocols occurs precisely when such a channel exists. The introduction of an infinity twistor does not create this channel.  Instead, it endows the channel with a spacetime interpretation, allowing its strength to be expressed geometrically in terms of spatial separation.  In this sense, QGEM experiments probe the existence of a bilocal quantum interaction which generates measurable entanglement, whereas the Newtonian $1/r$ dependence reflects a further assumption about the geometric content of that interaction.  Spacetime locality (in the sense of interactions represented via a distance-dependent metric) is therefore not a prerequisite for gravitationally mediated entanglement, but an emergent description of an underlying bilocal quantum structure.

\section*{Conflict of interests}

\noindent
The author has no conflicting interests to declare.

\section*{Data Availability Statement}

\noindent
No new data were generated during this study.

\end{document}